
%
%

\input phyzzx

\pubnum={Shiga-92-1}
\date={May, 1992}

\titlepage
\title{\seventeenrm
       Extended form method of antifield-BRST formalism for
       {\seventeenit BF} theories
      }

\author{Hitoshi IKEMORI\foot{E-mail address: ikemori@jpnyitp.bitnet}}
\address{Faculty of Economics, Shiga University
 \nextline
1-1-1 Bamba, Hikone, Shiga 522, Japan}

\abstract
{
The Batalin-Vilkovisky antifield action  for the
$BF$ theories
is constructed by means of the extended form method.
The BRST invariant BV antifield action
is directly
written down by making use of the extended forms
that involve all the required  ghosts and antifields.
 }

\def\tilde{\widetilde}

The $BF$ theory is a kind of the topological quantum field theories
(TQFT's),
which can be thought as a generalization of the Chern-Simons theory
to arbitrary dimensions [1-4].
Though they are related to the generalized linking numbers of the
extended objects, as far as the abelian $BF$ theories are concerned,
there are still some obstacles to proceed a quantization of
the non-abelian $BF$ theories in arbitrary dimensions ($\geq 4$).
The problems mainly come from the on-shell reducibility of
the symmetry in these systems.
It is well known that the Batalin-Vilkovisky (BV)
antifield-antibracket formalism
is a useful procedure
to construct a BRST invariant gauge fixed action
in a covariant manner [5].
Although the BV algorism is a sure method even for reducible systems,
one is required to do a tedious task to solve the BV master equation
especially for highly reducible theories.
There is known,
however, an algebraic method to obtain
the BRST algebra for the $BF$ theories
without solving directly the master equation [4].

 In this paper, we present a modified version of this algebraic method
to construct a BRST invariant BV action for BF theories,
which has obviously some advantages to the unmodified one.
The BV action that is a solution to the master equation
is explicitly written down
by means of the extended differential
forms that involves ``all'' the required  ghosts and antifields.
Though the BRST algebra is also derived from a simple condition,
the BV action itself can be obtained without referring
to a concrete form of the algebra.

Let us begin with a quick recollection of
the method of extended differential calculus on
the universal bundle [6-8]
in the case of topological Yang-Mills (TYM) theory.
The basic idea is an extension of the
exterior derivative to a sum of the usual exterior derivative $\ d\ $
and the BRST operator $\ s\ $,
$$
\tilde d = d + s \ .
 \eqno(1)
$$
According to this extension,
the differential form of type $(p,q)$ should be thought
as an extended form of total degree $ p+q $,
where we follow the convention
that a $p$-form with ghost number $q$ is called $(p,q)$-form
and we will often attach a subscript to the form indicating
these degrees as $\ \Phi _{(p,q)}\ $ when necessary.

The BRST algebra of TYM theory is obtained as follows.
The universal connection $\tilde A$ is considered as a sum of
the Yang-Mills connection $A$ and the Faddev-Popov ghost $c$ ,
$$
 \tilde A = A_{(1,0)} + c_{(0,1)}\ .
 \eqno(2)
$$
The curvature 2-form of $\tilde A$ whose elements we will denote as
$$
  \tilde F = F_{(2,0)} + \psi _{(1,1)} + \phi _{(0,2)}\ ,
 \eqno(3)
$$
is defined by
$$
 \tilde F := \tilde d\tilde A + {1\over 2}[\tilde A,\tilde A]\ ,
 \eqno(4)
$$
which is required to satisfy the Bianchi identity
$$
 \tilde D\tilde F = 0\ .
 \eqno(5)
$$
The explicit calculation of (4) and (5)
with the components (2) and (3) leads to the BRST algebra
$$
 \left\{
\eqalign
 {
   sA &= \psi - Dc          \cr
   sc &= \phi - {1\over 2}[c,c] \cr
   s\psi &= - D\phi -[c,\psi ]   \cr
   s\phi &= -[c,\phi ]          \ .\cr
 }\right.
 \eqno(6)
$$

 An algebraic method to obtain a BRST transformations for the $BF$
theories
has been presented in ref. [4],
which resembles the method above for the  TYM theory.
We will present a modified and extended version of this algebraic method.

The classical  invariant action  for the $BF$ theory in $D$-dimensional
spacetime has the form of
$$
  S = \int _{M_D}{\rm Tr} ( B \wedge F ) \ .
 \eqno(7)
$$
The fundamental fields in the action above are Lie algebra valued
1-form $A$ and $n$-form $B$,
where $\ F= dA + {1\over 2}[A,A]\ $ is
the curvature 2-form of connection $A$ and  $\ n=D-2,\ (D\geq 4)$.

The action has a symmetry with $(n-1)$-form $\varepsilon _{n-1}$ ;
$$
  \delta _{\varepsilon _{n-1}}A = 0\ ,
\quad \delta _{\varepsilon _{n-1}}B = D\varepsilon _{n-1}\ ,
 \eqno(8)
$$
as well as the ordinary Yang-Mills symmetry with $0$-form $\omega $ ;
$$
 \delta _\omega A = D\omega \ ,\quad \delta _\omega B = [\omega ,B]\ .
 \eqno(9)
$$

 The on-shell reducibility of the $\varepsilon $-symmetry
requires us to introduce a sequence of ghosts and ghosts-for-ghosts,
when we apply the BV algorism to this system.
We denote the sequence of ghosts descending from
$\ B=B_{(n,0)}\ $ as
$$
  B_{(n,0)}\rightarrow B_{(n-1,1)} \rightarrow
 \ldots \rightarrow B_{(n-q,q)}\rightarrow \ldots \rightarrow B_{(0,n)}\ .
 \eqno(10)
$$
There is of course, as required, an ordinary ghost $c$
for the Yang-Mills symmetry.

We propose that
the extended forms should be defined so as to include all
the required ghosts and BV antifields.
Therefore the extended forms $\tilde A$ and $\tilde B$ are defined by
$$
 \eqalignno
 {
  \tilde A   &= c_{(0,1)} + A_{(1,0)}
 + \sum _{q=0}^n B^*_{(2+q,-1-q)}    \ ,
        &(11{\rm a}) \cr
  \tilde B &= c^*_{(n+2,-2)} + A^*_{(n+1,-1)}
              + \sum _{q=0}^n B_{(n-q,q)}\ ,
        &(11{\rm b}) \cr
 }
$$
where $\ c^*_{(n+2,-2)}\ $, $\ A^*_{(n+1,-1)}\ $
and  $ B^*_{(2+q,-1-q)} $
are the BV antifield for
$\ c_{(0,1)}\ $, $\ A_{(1,0)}\ $ and $\ B_{(n-q,q)}\ $ respectively.

The conditions that lead to the BRST transformations are
$$
  \tilde F=0\ ,\quad \tilde D\tilde B=0\ ,
 \eqno(12{\rm a,b})
$$
which have the same form as the field equations
from the classical action (7)
except for  $\tilde {\ }\ $.
Expansion of the conditions (12a,b)
in the ghost number leads to the total BRST algebra
$$
 \left\{
 \eqalign
  {
    sc  \ \quad &= - {1\over 2}[c,c]
   \cr
    sA  \quad \quad &= - Dc
   \cr
    sB^*_{(2,-1)} \ &= - F - [ c, B^*_{(2,-1)} ]
   \cr
    sB^*_{(3,-2)} \quad &= - DB^*_{(2,-1)} - [ c, B^*_{(3,-2)} ]
   \cr
    sB^*_{(2+q,-q-1)}
   &=  - DB^*_{(1+q,-q)}
    -{1\over 2}\sum _{q'=0}^{q-2}
       [ B^*_{(2+q',-1-q')}, B^*_{(q-q',q'+1-q)}]
   \cr
   &\hskip120pt
    - [ c, B^*_{(2+q,-1-q)} ]\ ,\quad (2\leq q\leq n)\ ,
   \cr
 }\right.
 \eqno(13{\rm a})
$$
$$
 \left\{
 \eqalign
 {
   sc^* \quad
  &=   - DA^* - \sum _{q'=0}^{n} [ B^*_{(2+q',-1-q')}, B_{(n-q',q')} ]
           - [c, c^*]
  \cr
   sA^* \quad
   &=  - DB_{(n,0)}
       - \sum _{q'=0}^{n-1} [ B^*_{(2+q',-1-q')}, B_{(n-1-q',1+q')} ]
       - [c,A^*]
  \cr
   sB_{(n-q,q)}  &=  - DB_{(n-q-1,q+1)}
    - \sum _{q'=0}^{n-q-2} [ B^*_{(2+q',-1-q')}, B_{(n-q-2-q',q+2+q')} ]
   \cr
    &\hskip150pt - [c,B_{(n-q,q)} ] \ ,\quad ( 0\leq q\leq n-2)
  \cr
   sB_{(\ 1,n-1)} &=  - DB_{(0,n)} - [c,B_{(1,n-1)} ]
  \cr
   sB_{(\ 0,n)} \quad &=  -[c,B_{(0,n)} ]\quad .
  \cr
 }\right.
 \eqno(13{\rm b})
$$

 It should be remarked that our definition of the extended
forms (11a,b) is
enlarged from that of ref. [4] so as to include $A^*$ and $c^*$
together with $c$ .
They have employed a definition of the form
$$
 \tilde A = A_{(1,0)} + \sum _{q=0}^n B^*_{(2+q,-1-q)}\
,\quad \tilde B = \sum _{q=0}^n B_{(n-q,q)}\ ,
 \eqno(14{\rm a,b})
$$
in ref. [4] and have separately treated
the BRST algebra for the Yang-Mills symmetry.
It is due to the omission of $A^*$'s
that there is a restriction on ghost number
(${q\geq 1}$) in the expansion of their
conditions for BRST algebra,
$$
  \tilde F=0\ ,\quad \left( \tilde D\tilde B\right) _{q\geq 1}=0\ .
 \eqno(15{\rm a,b})
$$
Our definition of the extended forms seems to be more natural,
because the extension is a maximal one in $D$-dimensional spacetime
and there appear all the forms of degree $p=0$ to $D$
(once and for all).
The remarkable feature of our extended forms $\tilde A$
and $\tilde B$ is
that they look like antifields of each other, that is,
$\ {\tilde A}^* = \tilde B\ $ and $\ \tilde A = {\tilde B}^*\ $.

It seems also a remarkable advantage of our method that
a BRST invariant BV action is obtained quite easily and directly
by means of the extended forms as follows.
We point out that the BV antifield action is given by
$$
 {\cal S}_{\rm BV}= \int _{M_D}{\rm Tr}
 ( \tilde B\wedge \tilde F - \tilde B \wedge s\tilde A)\
 \eqno(16)
$$
in terms of the extended forms, which turns out to be equivalent to
$$
 \eqalign
 {
 {\cal S}_{\rm BV}=
\int _{M_D}\biggl[
  &\left( B_{(n,0)}\wedge F_{(2,0)}\right)
 + c^* \wedge {1\over 2}[c,c] + A^* \wedge Dc
  \cr
 &+ B_{(n,0)} \wedge [ c, B^*_{(2,-1)} ]
  + B_{(n-1,1)}\wedge
        \left( DB^*_{(2,-1)} + [ c, B^*_{(3,-2)} ]   \right)
  \cr
  &+ \sum _{q=2}^n B_{(n-q,q)}\wedge
  \cr
    \biggl(
    DB^*_{(1+q,-q)}
   &
      +{1\over 2}\sum _{q'=0}^{q-2}
      [ B^*_{(2+q',-1-q')}, B^*_{(q-q',q'+1-q)}]
    + [ c, B^*_{(2+q,-1-q)} ]
     \biggr)
  \biggr]\
  \cr
 }
 \eqno(17)
$$
in the component forms.
What we call BV action is a minimal solution to the master equation
and is considered as a generator for the BRST transformation
in the BV formalism.
It still contains antifields therefore is
to be gauge fixed by introducing a gauge fermion together with
antighosts and multipliers.

 The BV action (16) seems to have some relations to
the straightforward extension of
the classical invariant action (7),
$$
  \tilde S = \int _{M_D}{\rm Tr} ( \tilde B \wedge \tilde F ) \ .
 \eqno(18)
$$
It is obvious that the conditions (12a,b)
leading to the BRST algebra are nothing but formal field equations
from this extended action.
Though the extended action itself is not our object,
our BV action seems to be something like a Legendre transform of it.
Here we imply the Legendre transform of
$\ s\tilde A \ $ to $\ {\tilde A}^* = \tilde B$,
which is an analogy of translation from a Lagrangian to a Hamiltonian
provided that $ s\tilde A $ pretends to be a time derivative
of $\tilde A$,
though it is an odd time.

It may be understood as follows.
If we define odd ``canonical momenta'' $\tilde \pi _A$ and
$\tilde \pi _B$
of $\tilde A$ and $\tilde B$ as
$$
 \tilde \pi_A := {\partial \ \tilde {\cal L}\over \partial (s\tilde A)}
                = \tilde B \ ,\quad
 \tilde \pi_B := {\partial \ \tilde {\cal L}\over \partial (s\tilde B)}
                = 0   \ ,
 \eqno(19{\rm a,b})
$$
provided $\  \tilde S = \displaystyle\int _{M_D}\tilde {\cal L}\ $,
then it follows that ``Hamiltonian'' $\tilde {\cal H}$ is defined by
$$
  \tilde {\cal H}:= {\rm Tr}
     \left( \tilde \pi _A \wedge s\tilde A
      + \tilde \pi _B \wedge s\tilde B \right) - \tilde {\cal L}
       = {\rm Tr}\left( \tilde B\wedge s\tilde A
             - \tilde B \wedge \tilde F \right) \ .
 \eqno(20)
$$
Our BV action (16) is
nothing but this ``Hamiltonian'',
provided that its total sign is changed
and it is integrated on the manifold, that is,
$$
 {\cal S}_{\rm BV}= - \int _{M_D}\tilde {\cal H}\ .
 \eqno(21)
$$

It seems to be not an accident that the BV action can be written in
such a simple
form as (21) or (16).
A general structure  may be found  in our construction
which can be applicable to the other TQFT's.
There seems to be also a connection with
 the odd time canonical formulation of the BV formalism [9].
We will discuss these features
of our extended form method
in future publication [10].

\vskip20pt
\noindent{\fourteenbf Acknowledgement}

The author would like to thank S. Tsujimaru for fruitful discussions.
It is also a pleasure to thank S. Kitakado for reading
the manuscript and for encouragement.


\vskip30pt
\noindent{\fourteenbf References}

\item{[1]}
M. Blau and G. Thompson, Phys. Lett. B 228 (1989) 64.

\item{[2]}
G. Horowitz, Commun. Math. Phys. 125 (1989) 417.

\item{[3]}
D. Birmingham, M. Blau, M. Rakowski and G. Thompson,
Phys. Rep. 209 (1991) 129, and references therein.

\item{[4]}
J.C. Wallet, Phys. Lett. B 235 (1990) 71.

\item{[5]}
I.A. Batalin and G.A. Vilkovisky, Phys. Lett. B 102 (1981) 27,
\hfill\break
I.A. Batalin and G.A. Vilkovisky, Phys. Rev. D 28 (1983) 2567,
\hfill\break
I.A. Batalin and G.A. Vilkovisky, J. Math. Phys. 26 (1985) 172.

\item{[6]}
H. Kanno, Z. Phys. C 43 (1989) 477.

\item{[7]}
L. Baulieu and I.M. Singer, Nucl. Phys. B (Proc. Suppl.) 5B (1988) 12.

\item{[8]}
S. Ouvry, R. Stora and P. van Baal, Phys. Lett. B 220 (1989) 159.

\item{[9]}
\"O.F. Dayi, Mod. Phys. Lett. A 4 (1989) 361.

\item{[10]}
H. Ikemori, Shiga Univ. preprint, in preparation.

\end